\documentclass[11pt,a4paper]{article}
\pdfoutput=1
\usepackage{jcappub}
\usepackage{subfigure}
\usepackage{booktabs}
\usepackage{graphicx,longtable,natbib,mathrsfs,color}
\usepackage{txfonts}
\usepackage{threeparttable}

\newcommand{\apjl}{Astrophys. J. Lett.}
\newcommand{\apjs}{Astrophys. J. Suppl. Ser.}
\newcommand{\aj}{Astron. J.}
\newcommand{\mnras}{Mon. Not. R. Astron. Soc.}
\newcommand{\physrep}{Physics Letters. B}
\newcommand{\jcap}{J. Cosmology Astropart. Phys.}

\newcommand{\prd}{Physics Review D}
\newcommand{\apj}{Astrophys. J. }

\title{Breaking through the high redshift bottleneck of Observational Hubble parameter Data: The Sandage-Loeb signal Scheme}
\author[a]{Shuo~Yuan,}
\author[a]{Siqi~Liu,}
\author[a]{and Tong-Jie~Zhang}

\affiliation[a]{Department of Astronomy, Beijing Normal University,
Beijing, 100875, China}
\emailAdd{yuanshuoastro@pku.edu.cn}\emailAdd{siqiliu@mail.bnu.edu.cn} \emailAdd{tjzhang@bnu.edu.cn}

\abstract{ We propose a robust scheme to measure the Hubble parameter $H(z)$ at high redshifts by detecting the Sandage-Loeb signal (SL signal) which can be realized by 
the next generation extremely large telescope. It will largely extend the current observational Hubble parameter data (OHD) 
towards the redshift range of $z \in [2.0,5.0]$ where other 
dark energy probes is difficult to provide useful information of the cosmic expansion. To quantify the capability of such future measurement to constrain cosmological models, we simulate observational data for a CODEX (COsmic Dynamics and EXo-earth experiment)-like survey. We find that the SL signal scheme brings the redshift upper-limit of OHD from $z_\mathrm{max}=2.3$ to $z_\mathrm{max}\simeq 5.0$,   
provides more accurate constraints on different dark energy models, and greatly changes the degeneracy direction of the parameters.
For the $\Lambda$CDM case, the accuracy of $\Omega_m$ is improved by $58\%$ and the degeneracy between $\Omega_m$ and $\Omega_
{\Lambda}$ is rotated to the vertical direction of $\Omega_m -\Omega_\Lambda$ plane; for the $w$CDM case, the accuracy  of $w$ is improved by $15\%$. The Fisher matrix forecast on different time-dependent $w(z)$ cosmological model is also performed. }

\arxivnumber{1311.1583}
\begin{document}

\maketitle
\flushbottom

\section{Introduction}\label{sec:intro}
The concept that the universe is undergoing an accelerated expansion
has been strongly proved. One of the popular methods is the geometrical probe that is based on distance measurement: the luminosity distance measurement with the type  \uppercase \expandafter {\romannumeral 1}a supernova \citep{1998AJ....116.1009R,1999ApJ...517..565P} (``the standard candle") and the angular diameter distance of the first CMB acoustic peak, Baryon Acoustic Oscillations (BAO)\citep{2005ApJ...633..560E} (``the standard ruler").
\medskip{}

It should be noted that the   experiments mentioned above are based on observations of different objects at different distances. Furthermore, they require to invoke the Copernican cosmological principle and Einstein's equations of motion. Therefore, measuring the dynamics of the universe in a direct and straightforward way will provide be a new fundamental method to measure the Hubble parameter $H(z)$.
\medskip{}

The determination of $H(z)$ is directly related to the expansion history of the universe by its definition: $H = 
\dot{a}/{a}$, where $a$ denotes the cosmic scale factor and $\dot{a}$ is its 
rate of change with respect to the cosmic time. Apart from depicting the accelerated expansion of the universe, 
the Hubble parameter provides us an alternative probe to study different cosmology models, to investigate new parameters and to constrain some key stages of the expansion history. 
The latest set of the Observational Hubble parameter Data (OHD) has been published in a sample of 28 measurements within $0.07\leqslant z\leqslant2.30$ \citep{2013ApJ...766L...7F}, via the cosmic chronometers \citep{2005PhRvD..71l3001S,2010ApJS..188..280S,2012JCAP...08..006M,
2014RAA....14.1221Z} and the BAO peak approaches\citep{2009MNRAS.399.1663G,2012MNRAS.425..405B}.  
 The applications of available OHD and its potential in constraining cosmological parameters is explored in \cite{2009MPLA...24.1699L} and \cite{2006ApJ...650L...5S}. Ref. \cite{2011ApJ...730...74M} analyzed the simulated OHD in $0.1 \leqslant z \leqslant 2.0$ and pointed out that the OHD could be a powerful probe in future experiments. Recently, the nonparametric reconstruction of dynamical dark energy using OHD %\citep{2013arXiv1310.0870Y} 
\citep{2013PhRvD..88j3528Y} was accomplished. 
However, getting beyond $z=2.3$ represents a bottleneck with current techniques. 
Can we make a dynamical scheme, or the so-called ``real-time cosmology" method \citep{2012PhR...521...95Q}, to measure the $H(z)$ at higher redshifts?
\medskip{}

Sandage studied
a possible cosmological tool to directly measure the temporal variation due to the redshift of extra-galactical sources\cite{1962ApJ...136..319S}.
Loeb revisited this idea and pointed out
that the spectroscopic techniques developed for detecting the reflex motion of stars induced by unseen orbiting planets could be used for
detecting the redshift variation of QSO Lyman$-\alpha$ absorption lines \citep{1998ApJ...499L.111L}, i.e., the so-called ``Sandage-Loeb" (SL) effect. The QSOs used in this method lie within the redshift range of $2.0\lesssim z\lesssim5.0$. The SL effect has been widely adopted in various cosmological researches. 
It was first employed to explore the dark energy redshift desert in \citep{2007PhRvD..75f2001C}. This scheme is extended to constrain other dark energy models: interacting dark energy\cite{2007MNRAS.382.1623B}, Chaplygin gas\cite{2012PhRvD..86l3001M}, the new agegraphic and Ricci dark energy models\cite{2010PhLB..691...11Z} and some modified gravity theories \cite{2010PhLB..692..219J}  \cite{2013PhRvD..88b3003L}. These previous works used the simulated redshift drift data in the redshift range $2.0 \leqslant z \leqslant 5.0   $ to investigate the expected constrains for cosmological models. 
\medskip{}  

The forthcoming major observation facilities, e.g. the European Extremely Large Telescope (E-ELT)\footnote{\url{http://www.eso.org/public/teles-instr/e-elt/}}, will offer a stable and higher resolution spectroscopic
detection, which allows to measure the Sandage-Loeb signal (SL signal or the so-called ``redshft drift").
The undergoing project CODEX (COsmic Dynamics and EXo-earth experiment)\footnote{\url{http://www.iac.es/proyecto/codex/}} aims to detect the SL signal with E-ELT.  
\medskip{}

In this paper, we introduce the theory of SL scheme and present
the strategy to measure high redshift OHD from a CODEX-like survey in section \ref{sec:SL signal}. In section \ref{sec:lcdm} and section \ref{sec:fisher}, using a Monte Carlo Markov Chain (MCMC) approach, we also perform a test to explore the potential power of our new strategy to constrain the cosmological  parameters of future OHD from a CODEX-like survey, in various dark energy models.  We use the publicly available code \texttt{PyMC} \footnote{\url{https://github.com/pymc-devs/pymc}} to perform a full MCMC analysis. In the end we draw our conclusions in section \ref{sec:conclu}.

\section{From Sandage-Loeb signal to OHD: Theory and Data}\label{sec:SL signal}
\subsection*{The SL signal $\dot{v}$}
The possibility of direct and completely model independent measurements of the redshift drift due to the expansion of the universe was first studied by Sandage 
(for detailed discussions see \cite{1962ApJ...136..334M}).
%In a homogeneous and isotropic universe, 
 Here we show SL signal, or redshift drift\footnote{Strictly speaking, the term ``redshift drift" is referred to the $ \Delta z / \Delta t_0$ while the $\Delta v / \Delta t_0$ called ``Sandage-Loeb signal(SL signal)" in this letter. In fact these two qualities are equivalent and can be converted to $H(z)$(see Eq.(\ref{conv})). }, as 
\begin{equation}
\label{vdot}
\dot{v}\equiv\frac{\Delta v}{\Delta t_{0}}=\frac{cH_{0}}{1+z}\biggl[1+z-\frac{H(z)}{H_{0}}\biggr],
\end{equation}
where $\dot{v}$ is the SL signal which can be measured by measuring the Lyman-$\alpha$ absorption system in the QSOs' spectrum
for a decade of $\Delta t_0$.
\medskip{}

The undergoing CODEX project is fed by E-ELT. One of its science targets is to directly measure the accelerating expansion of the universe by 
detecting the cosmological redshift drift of the Lyman-$\alpha$ forest from QSO lying in $2.0 \lesssim z \lesssim 5.0$.
Ref. \cite{2008MNRAS.386.1192L} estimates the statistical errors by performing the Monte Carlo simulations of Lyman-$\alpha$
absorption lines. The uncertainty on $\Delta v$ can be written as
\begin{equation}
\label{eq2}
\sigma_{v}=
\begin{cases}
1.4\Bigl(\frac{S/N}{2350}\Bigr)^{-1}\Bigl(\frac{N_{\mathrm{QSO}}}{30}\Bigr)^{-0.5}\Bigl(\frac{1+z_{\mathrm{QSO}}}{5}\Bigr)^{-1.7}\Bigl[\mathsf{\frac{\mathsf{cm}}{s}}\Bigr],z_{\mathrm{QSO}}\leq4;\\
1.4\Bigl(\frac{S/N}{2350}\Bigr)^{-1}\Bigl(\frac{N_{\mathrm{QSO}}}{30}\Bigr)^{-0.5}\Bigl(\frac{1+z_\mathrm{QSO}}{5}\Bigr)^{-0.9}\Bigl[\mathsf{\frac{\mathsf{cm}}{s}}\Bigr],z_\mathrm{QSO}>4,
\end{cases}
\end{equation}
where $S/N$ is the spectral signal-to-noise defined per 0.00125 nm pixel, $N_{\mathrm{QSO}}$
is the number of QSOs and $z_{\mathrm{QSO}}$ is the quasar's redshift.
\medskip{}

\subsection*{The ``SL signal scheme": from $\dot{v}$ to $H(z)$}
Suppose we detect the SL signal $\dot{v}$ during $\Delta t_0$, 
we can convert it to the $H(z)$ and its error $\sigma_{H(z)}$ via:
\begin{equation}
\label{conv}
\begin{cases}
H(z)=(H_0-\dot{v} / c )(1+z)\\
\sigma_{H(z)}=(1+z)\sigma_v /( c\Delta t_0 ).
\end{cases}
\end{equation}
Therefore, the SL signal scheme can be used to extend the current OHD to a higher range of $2.0 \lesssim z \lesssim 5.0$. This is a new method to measure the high redshift $H(z)$, and here we call it ``SL signal scheme".
Note that the latest OHD contains 28 points and covers the region from $z=0.1$ to $2.3$. The method described above can be used to complement the current OHD at higher redshifts.
\medskip{}

\begin{figure}
    \includegraphics[width=1.0\textwidth]{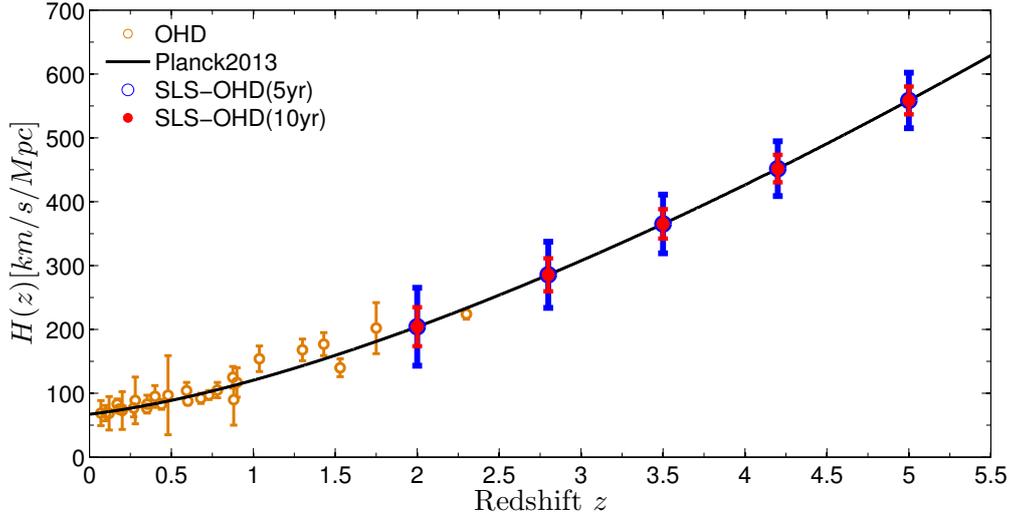}
    \label{fig1}
    \caption{The forecast Sandage-Loeb signal OHD (``SLS-OHD") and the current OHD. The brown dots are the current available OHD and red filled circle represents the forecast data from a observation interval of 10 years, while blue ones from 5 years. The significant decrease of error bars with $\Delta t_0$ is clearly seen.}
\end{figure}

\subsection*{Data}
In the following analysis, we use the data model in Eq.(\ref{conv}) to make a forecast  for a SL signal experiment on
a CODEX-like survey. Namely, we consider a survey observing a total  number of $150$ QSOs uniformly distributed in redshift bins of  $z_\mathrm{QSO}=[2.0,2.8,3.5,4.2,5.0]$
with $S/N$ of $3000$ 
%Here we refer the number QSO reported in \cite{2010PhLB..691...11Z} and \citep{2007PhRvD..75f2001C}, both of them considering  $N_{\rm{QSO}}=240$)
, by setting throughout this paper $N_{\rm{QSO}} =30$ per redshift bin, i.e., 150 QSO in total. 
We consider a fiducial concordance $\Lambda$CDM model to generate the mock data (the parameters have been set to the latest values obtained from the Planck analysis \citep{2013arXiv1303.5076P}), Then we use the mock data to examine the capability of the future OHD and check the improvement
against current OHD. Both the forecast data and the current OHD are shown in Fig.1. 
\medskip{}

We have 2 ways to generate the simulated $H(z)$ data using SL signal scheme.
One way is to set the simulated data ``directly" on the fiducial model like in the way used in \citep{2007PhRvD..75f2001C} and \cite{2012PhRvD..86l3001M}. This is only available under an ideal condition in which the observational values are not shifted away from the fiducial model predictions.
Another way is to generate the mock data ``shifted" from fiducial model by an offset $\delta(z)$:
\begin{equation}
\label{bsl}
H(z) = H^{\star}(z) + \delta(z).
\end{equation}
where $H^{\star}(z)$ is just the fiducial model value.
The offsets $\delta(z) $ are stochastic variables from  a Gaussian distribution
\begin{equation}
\delta(z) \sim \mathcal{N}(0, \sigma_{H(z)}).
\end{equation}
We choose the later one as the scheme to generate the mock data set ``SL-OHD".
\medskip{}

\begin{table}[!h]
\label{epsl_L_ohd_ohd+}
\tabcolsep 0pt
\vspace*{-10pt}
\begin{center}
\def\temptablewidth{0.8\textwidth}
{\rule{\temptablewidth}{1pt}}
\begin{tabular*}{\temptablewidth}{@{\extracolsep{\fill}}cc}
Data Set Name & Descriptions                                                   \\ \midrule
OHD           & The available OHD data (28 points).                             \\
SL-OHD       & The mock Sandage-Loeb signal $H(z)$ with offset $\delta(z)$.           \\
EOHD          & The ``Extended OHD": the combination set of OHD and SL-OHD.                        \\ \bottomrule
\end{tabular*}
       {\rule{\temptablewidth}{1pt}}
\end{center}
	   \vspace*{-18pt}
       \caption{ The descriptions of data sets used in this paper. }
\end{table}

The summary of mock data sets in this paper are listed in Table {1}. The comparisons of constraining results are between OHD and EOHD. The duration of the SL experiment is $\Delta t=10$ yrs if not differently specified. 

\section{Test on Dark Energy Models}\label{sec:lcdm}
In this section, we focus on the potential improvement of high redshift OHD using the SL measurement strategy described in Eq.(\ref{conv}).
Because of the possible non-gaussian posteriors in   
 parameter space, we should perform a full MCMC analysis to explore the posterior distribution. Here we use the MCMC code \texttt{PyMC} \citep{Patil:Huard:Fonnesbeck:2010:JSSOBK:v35i04} to sample the parameter space.  
\medskip{}

It has been shown in Eq.(\ref{conv}) that $H(z)$ and $\dot{v}$ are equivalent. However, there are some reasons and advantages that make the conversion of $H(z)$ from $\dot{v}$  meaningful:
\begin{itemize}
\item The ``SL signal scheme" we present in Sec.2 is a complete model independent measurement of $H(z)$. Because the SL signal is direct and independent against any pre-assumed model, $H(z)$ extracted through SL scheme is a straightforward and directly comparable with the current 28 data points  from "cosmic chronometers".
\item The overlapped redshift interval (around $z \simeq 2$) between current OHD and SL scheme OHD can be used for self-calibration for the $H(z)$ values. The available OHD can be obtained up to $z_{max}=2.3$, while the SL signal scheme can present a minimal redshift around 2. Hence the $H(z)$ values retrieved from different methods will calibrate each other and enhance the accuracy at redshift around $2$.  
\item The current OHD observational methods are not efficient on derivating   the values of $H(z)$ in the redshift range of $2.0 \leq z \leq 5.0 $: only one point at $z=2.3$  is obtained so far.  
 However, the SL signal scheme should be able to provide some $H(z)$ values in this blank and break through the redshift bottleneck of OHD. 
\end{itemize}

In this section we focus on the improvement of OHD on constraining the cosmological models after the introduction of the additional points obtained with the  SL signal scheme. Namely, we compare the OHD with the EOHD (see Tab.1). We use OHD and EOHD to fit  $\Lambda$CDM and $w$CDM models. 
For each model, we run MCMC sampling in its parameter space, finding the best fit values, confidence  regions  and convariance matrix. After we get these fitting quantities, the comparison between them should tell us the improvements and the capability of the SL signal scheme. They should be investigated from the following 3 aspects of statistical quantities from Markov chains: a) How much the errorbars can be reduced with the mock data? b) How does the Figure of Merit (FoM) changed? and c) What is the difference between the degeneracy of parameters of the mock data and OHD? 
The details about these aspects are explained in the following. 
%+++ adding 3 def. here!
\paragraph*{The relative accuracy improvement quality $\varepsilon$:}
Suppose the current data can give marginalized  1-$\sigma$ probability  interval of the model parameter (the ``error bar") $\sigma_0(\theta_i)$ and our forecast data can give the new $\sigma_1(\theta_i)$ where $\theta_i$ is the $i-$th parameter of a given model. We can define the relative accuracy improvement quality $\varepsilon$:
\begin{equation}
\label{31}
\varepsilon(\theta_i) = \frac{\sigma_0(\theta_i) - \sigma_1(\theta_i)}{ \sigma_0(\theta_i) }.
\end{equation}

Apparently, the larger positive $\varepsilon(\theta_i)$ the better the forecast data performs in constraining $(\theta_i)$. We use this statistical  quality to examine the improvement of the forecast data on a given parameter constraining.

\paragraph*{The change of Figure of Merit (FoM) $\Delta FoM$:}

We follow the definition of DEFT (Dark Energy Task Force)\citep{2006astro.ph..9591A} on the Figure of Merit (FoM):
\begin{equation}
\label{fom}
FoM = \frac{\pi}{A} = \frac{1}{\sigma(\theta_i)\sigma(\theta_j)\sqrt{1-\rho_{ij}}},
\end{equation}
where $A$ is the area of 1-$\sigma$ confidence regions (or the so-called "confidence ellipses") of parameter $\theta_i$ and $\theta_j$, $\sigma(\theta_i)$ is the 1-$\sigma$ confidential interval of parameter $\theta_i$ and $\rho_{ij}$ is the correlation coefficient between parameters $\theta_i$ and $\theta_j$. The smaller $FoM$ is, the better constraint we get. For the current data, we can define the $FoM_0$ and for the forecast data we have $FoM_1$. %meanwhile the footnotes of variables in the rightest side of Eq.(\ref{fom}) should be 0 and 1 for the corresponding data. 
Therefore, the quantity $\Delta$FoM, defined as
\begin{equation}
\label{dfom}
\Delta FoM = \frac{FoM_1 - FoM_0}{FoM_0}
\end{equation} 
tells us how the constraint from the forecast data is better than the current data in percentage.

\paragraph*{The change of parameter degeneracy:}
The degeneracy direction between $\theta_i$ and $\theta_j$ is determined by $\rho_{ij}$. Namely, we consider the covariance matrix $\boldsymbol{C}$ of the parameter space $\{ \boldsymbol{\theta} \}$:
\begin{equation}
\label{cov}
\boldsymbol{C}_{ij} = \sigma(\theta_i) \sigma(\theta_i) \rho_{ij},
\end{equation} 
where the correlation coefficient $\rho_{ij} \equiv 1$ if $i=j$. The covariance matrix $\boldsymbol{C}$ can be estimated from the MCMCs. The difference between $\rho_{ij}$ from the current and the forecast data tells us the change of the change of the degeneracy.

\subsection{Standard Cosmological Model}

In the $\Lambda$CDM scenario,  dark energy is
 constant in time, with an equation of state parameter $w=p/\rho\equiv-1.$ 
 According to the Friedmann equation, the prediction of Hubble parameter is
\begin{equation}
\label{hz}
H^{\star}(z)=H_{0}\sqrt{\Omega_{m}(1+z)^{3}+\Omega_{\Lambda}+\Omega_{k}(1+z)^{2}},
\end{equation}
where $\Omega_{m}+\Omega_{\Lambda}+\Omega_{k}=1$. 
We perform the $\chi^{2}$-statistics for this model:
\begin{equation}
\label{chi}
\chi_\mathrm{OHD}^{2}(\mathcal{D},\boldsymbol{\Omega})=\sum_{i=1}^{N}\biggl(\frac{H^\mathrm{obs}(z_{i})-H^{\star}(z_{i}|\boldsymbol{\Omega})}{\sigma_{i}}\biggr)^{2},
\end{equation}
where $\mathcal{D}$ is the observational Hubble parameter data in the form of $\{z_{i},H^\mathrm{obs}(z_{i}),\sigma_{i}\}$ and $H^{\star}$ is the theoretical prediction of Hubble parameters given by the cosmology.
Therefore the model's free-parameter vector is $\boldsymbol{\Omega}=(H_{0},\Omega_{m},\Omega_{\Lambda}).$
\medskip{}

The capability of OHD and EOHD to constrain this model is shown in  Fig.2 . We find that  EOHD changes the direction of degeneracy significantly towards the vertical direction of the  $\Omega_m-\Omega_\Lambda$  plane compared with the OHD which provides a better constraint. From the right panel of Fig.2, we find EOHD breaks the current degeneracy between $H_0$ and $\Omega_m$ and produces a tighter constraints.
\medskip{}

To illustrate the improvement of new data on constraining $\Lambda$CDM quantitatively, we compare the marginalized probability distribution function of model parameters using Eq.(\ref{epsl_L_ohd_ohd+}). 
We find that in the case of $\Lambda$CDM, the EOHD can improve all of the parameters with $\varepsilon$ varies from 13.5\% (for parameter $H_{0}$) to 58.5\% (for parameter $\Omega_{m}$).
\medskip{}
%------------
\begin{table}[!h]
\label{epsl_L_ohd_ohd+}
\tabcolsep 0pt
\vspace*{-10pt}
\begin{center}
\def\temptablewidth{0.8\textwidth}
{\rule{\temptablewidth}{1pt}}
\begin{tabular*}{\temptablewidth}{@{\extracolsep{\fill}}ccccc}
Model & $\sigma(\boldsymbol{\Omega})$ & OHD & EOHD & $\varepsilon$(\%)     \tabularnewline  
\hline 
$\Lambda$CDM & $\sigma(H_{0})$             & 4.45 & 3.85 & 13.5   \tabularnewline
             & $\sigma(\Omega_{m})$        & 0.08 & 0.03 & 58.5   \tabularnewline 
             & $\sigma(\Omega_{\Lambda})$  & 0.25 & 0.15 & 40.0    \tabularnewline
             & $\sigma(\Omega_{k})$        & 0.32 & 0.18 & 43.7    \tabularnewline
\hline 
\end{tabular*}
       {\rule{\temptablewidth}{1pt}}
\end{center}
	   \vspace*{-18pt}
       \caption{The improvement of EOHD on 1-$\sigma$ error on cosmological  parameters in the case of $\Lambda$CDM. Here $\boldsymbol{\Omega}$ represents the model parameter space. The column of $\varepsilon$ is the relative improvement ratio for each model parameter defined in Eq.(\ref{31}). }
\end{table}
%------------
%% fig.2
\begin{figure}
\begin{minipage}[t]{0.5\linewidth}
\centering
\includegraphics[width=3.3in]{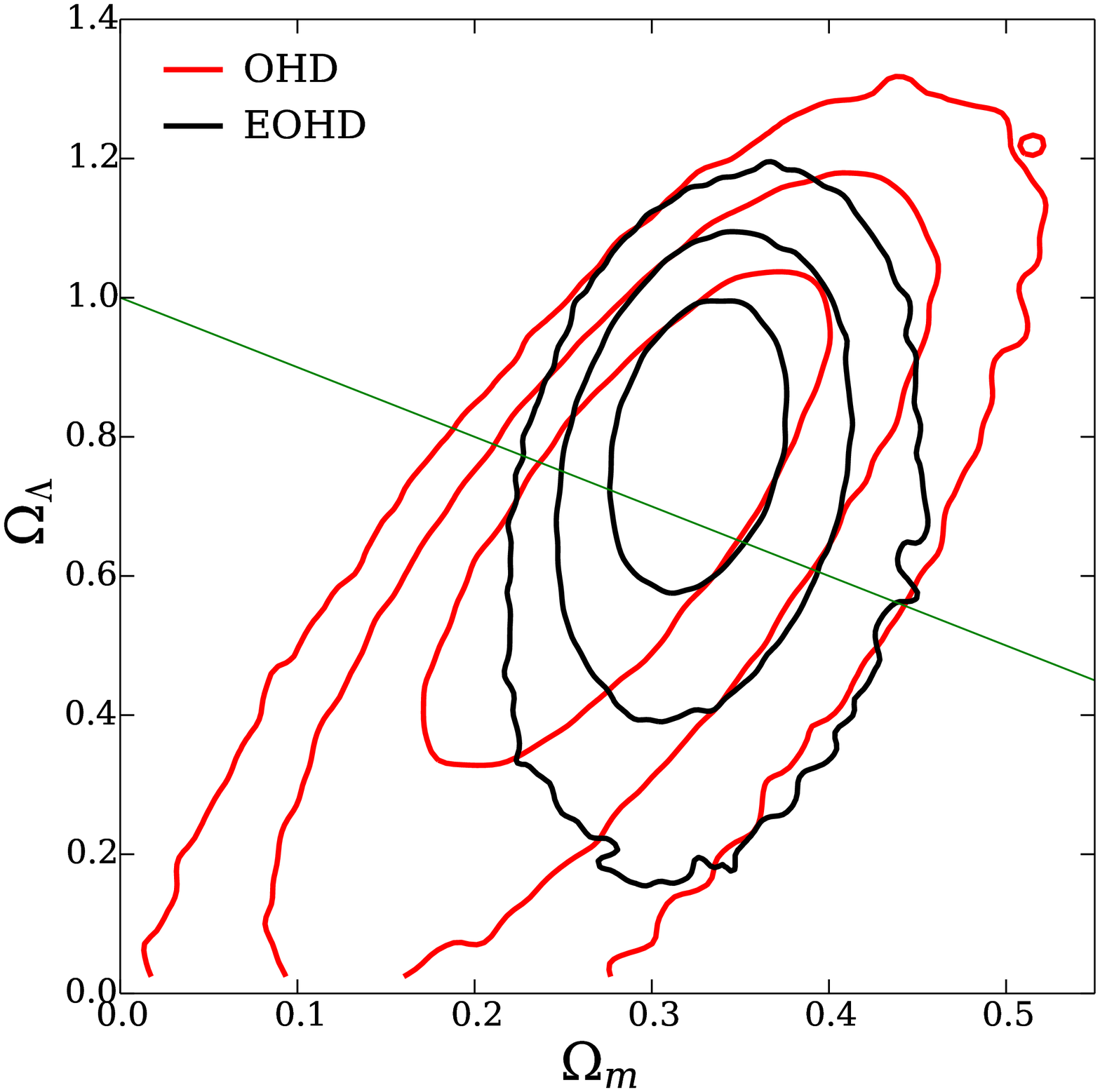}
%\caption{fig1}
\label{fig:side:a}
\end{minipage}%
\begin{minipage}[t]{0.5\linewidth}
\centering
\includegraphics[width=3.3in]{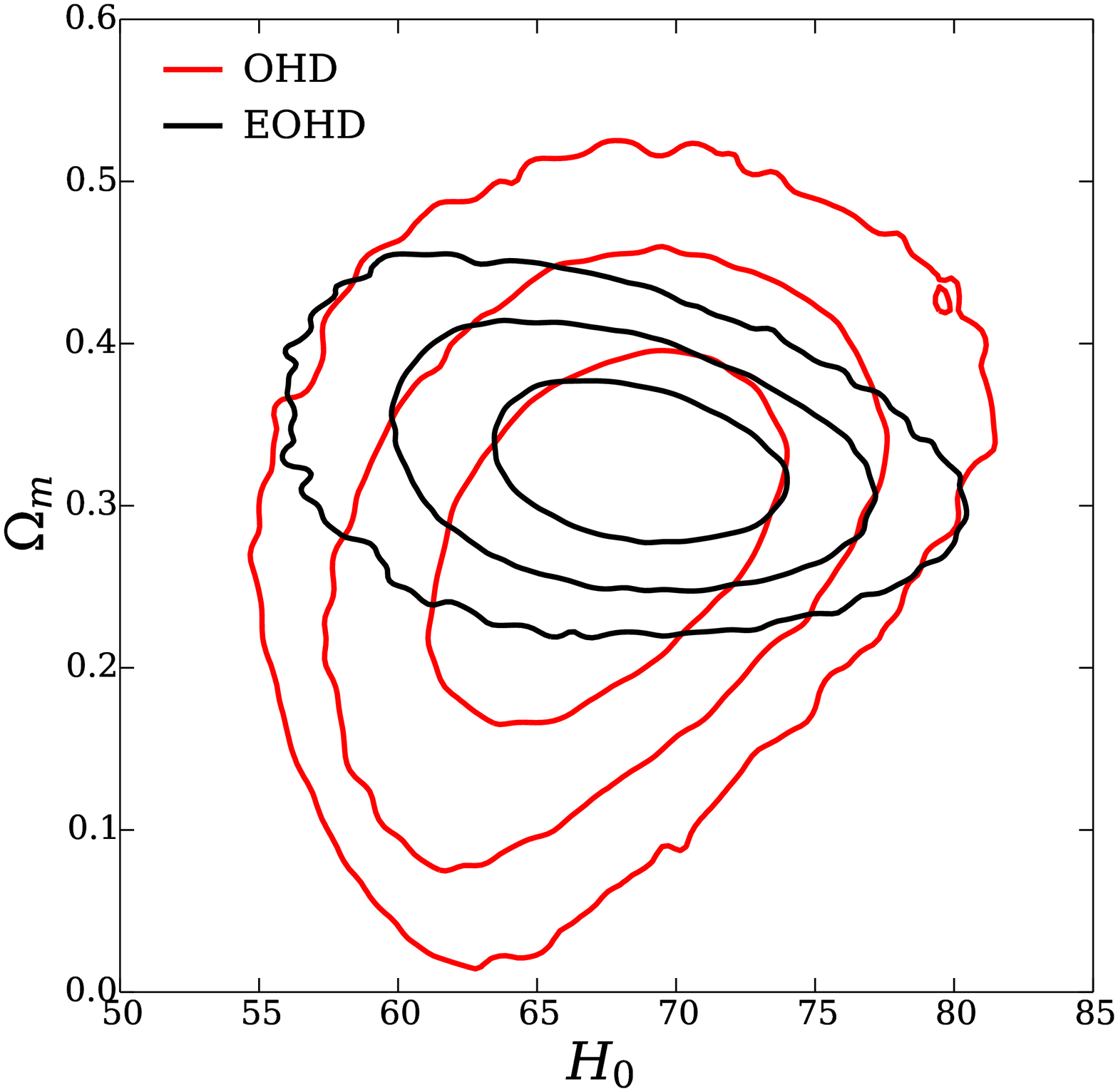}
%\caption{fig2}
\label{fig:side:b}
\end{minipage}
\caption{Constraints on $\Lambda$CDM using OHD (red contours) and EOHD(black contours). \textit{Left:} Marginalized 2-dimensional probability distribution function for $\Omega_{m}$and $\Omega_{\Lambda}$. The green line is the case of $\Omega_{k}=0$, the flat $\Lambda$CDM, i.e. $\Omega_{\Lambda}+\Omega_m=1 $. Compared with the current OHD, the new set will change the degeneracy direction significantly towards the vertical direction the $\Omega_m-\Omega_{\Lambda}$ plane and produce a tighter constraint. \textit{Right:}   Marginalized 2-dimensional probability distribution function for $\Omega_{m}$and $H_{0}$. Compared with the current OHD, the new sets EOHD breaks the degeneracy direction and produces a tighter constraint.}
\end{figure}

We can obtain the covariance matrix $\boldsymbol{C}$ from the MCMC samples and extract the coefficient matrix $\boldsymbol{R}_{ij} = \rho_{ij}$ which shows the degeneracy between parameters. The results are shown in Fig.(3) and $\Delta FOM$ is shown in Table 3. 
\medskip{}
%------
%------------
\begin{table}[!h]
\label{fom_L_ohd_ohd+}
\tabcolsep 0pt
\vspace*{-10pt}
\begin{center}
\def\temptablewidth{0.8\textwidth}
{\rule{\temptablewidth}{1pt}}
\begin{tabular*}{\temptablewidth}{@{\extracolsep{\fill}}ccccc}
Model & $FoM(\boldsymbol{\Omega}_i,\boldsymbol{\Omega}_j) (i \neq j)$ & OHD & EOHD & $\Delta FoM$(\%)  \tabularnewline
\hline 
$\Lambda$CDM & $FoM(\Omega_{m},\Omega_{\Lambda})$ 			& 128.64 & 274 & 112.59	  	\tabularnewline
 			 & $FoM(H_0,\Omega_m)$  	                        	& 4.32   & 8.13& 88.66	  	\tabularnewline 
\hline 
\end{tabular*}
       {\rule{\temptablewidth}{1pt}}
\end{center}
	   \vspace*{-18pt}
       \caption{The improvement of EOHD on FoM in the case of $\Lambda$CDM. Here $\boldsymbol{\Omega}$ represents the model parameter space. The definition of $\Delta FoM$ is in Eq.(\ref{fom}).} %The negative larger $\Delta FoM$ is, the 
\end{table}
%------------
%------

\begin{figure}[!h]
\label{cov_L}
\begin{minipage}[t]{0.5\linewidth}
\centering

\includegraphics[width=2.9in]{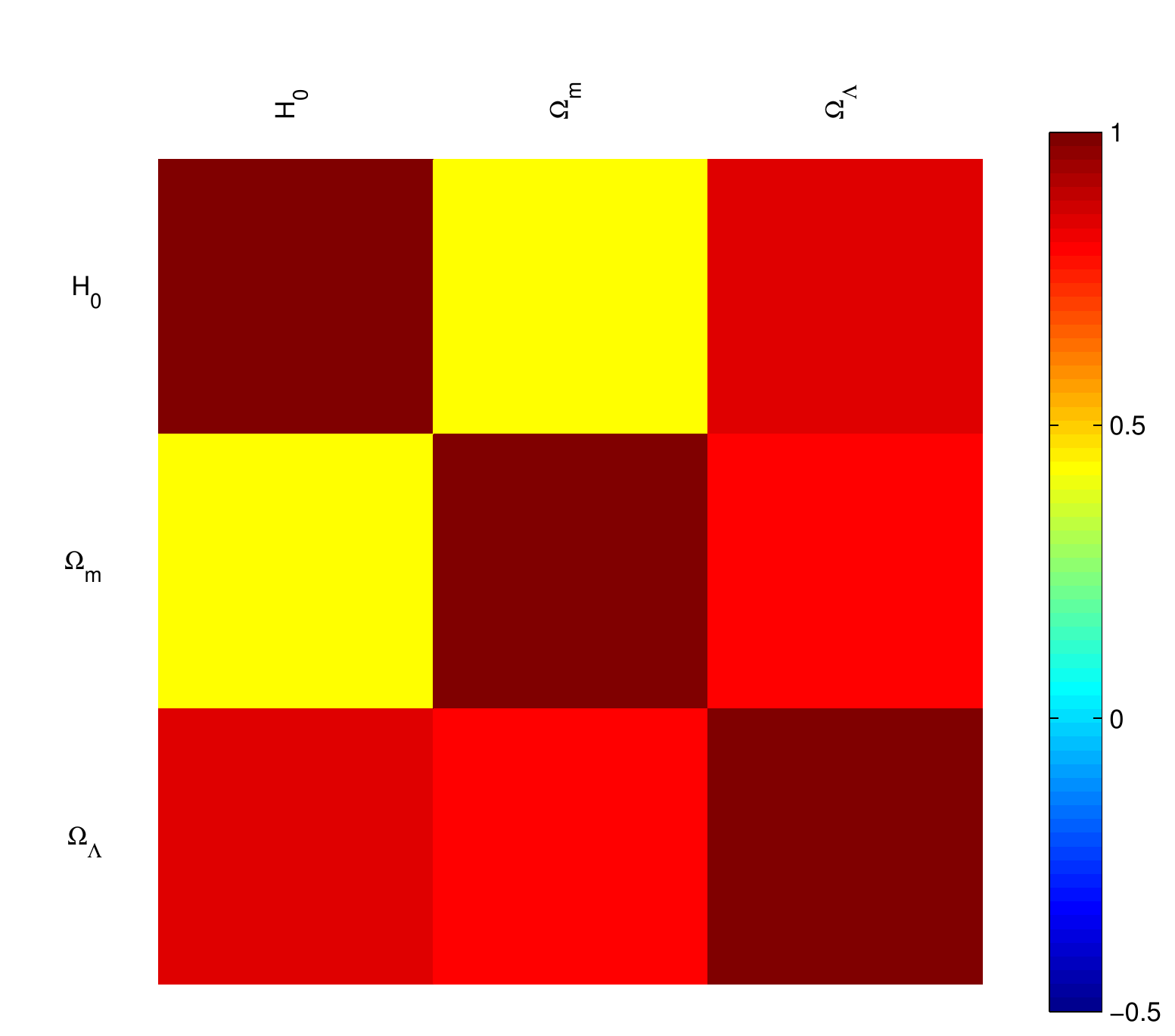}
%\caption{fig1}
\label{fig:side:a}
\end{minipage}%
\begin{minipage}[t]{0.5\linewidth}
\centering
\includegraphics[width=2.9in]{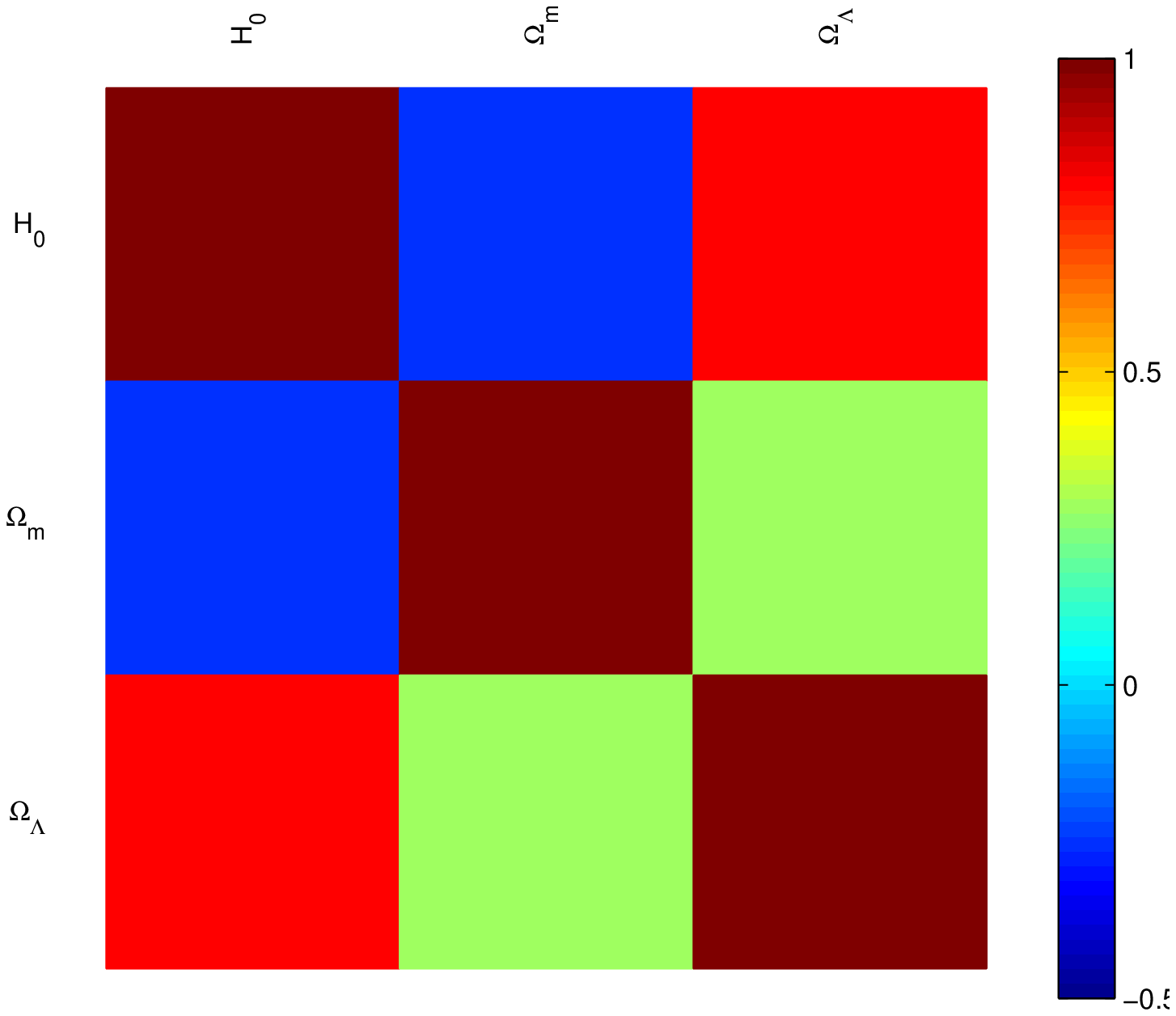}
%\caption{fig2}
\label{fig:side:b}
\end{minipage}
\caption{The correlation matrix $\rho_{ij}$ of $\Lambda$CDM. The left side is using OHD and the right one is using EOHD. The two color bars share the same scale. It is clear that the degeneracy is changed.}
\end{figure}

The comparisons just discussed imply that the 5 new mock $H(z)$ data in EOHD obtained with the SL signal scheme can help to significantly improve the constraint on cosmological parameters.

\subsection{$w$CDM Scenario}
\label{sec:nonstdde}
%\subsubsection*{OHD vs OHD$^+$}
Another widely studied model of dark energy is $w$CDM parameterization where the dark energy is characterized by a constant equation of state parameter $w=p/\rho$ not necessarily equal to -1. The Friedmann equation in this model is:
\begin{equation}
\label{hz2}
\Bigl[\frac{H^{\star}(z)}{H_{0}}\Bigr]^{2}=\Omega_{m}(1+z)^{3}+\Omega_\mathrm{DE}(1+z)^{3(1+w)}.
\end{equation}
Here a flat universe is assumed i.e., $\Omega_{m}+\Omega_\mathrm{DE}=1.$
The same $\chi^{2}-$statistics is performed using Eq.(\ref{chi}) 
with the free-parameter vector $\boldsymbol{\Omega}=(H_{0},\Omega_{m},w)$. The numerical results demonstrating the capability of OHD and EOHD to constrain this model are showed in Fig.4. 
We find that the contours in the $\Omega_m$-$w$ plane and in the $H_0 - \Omega_m$ plane presents strong degeneracies but the EOHD present a narrower and more elliptical region. Therefore the improvement of constraining is significant in $\Delta FOM$ (Table 5.) but weak in marginalized errorbars listed in Table 4: the improvement on $w$ is big ($\varepsilon=14.28\%$) but not better on $\Omega_m$.
\medskip{}
 
 The comparison between OHD and EOHD under the $w$CDM implied that the SL signal scheme do improve the quality of OHD and reduce the errorbars of parameters. The high redshift OHD have tiny effect on the $H(z=0)=H_0$, so the $\varepsilon^R$ of $H_0$ is negligible. The quantitative description of the change of degeneracy is shown in Fig.(5).
\medskip{}
\begin{figure}[h]
\label{cov_L_ellp}
\begin{minipage}[t]{0.5\linewidth}
\centering
\includegraphics[width=3in]{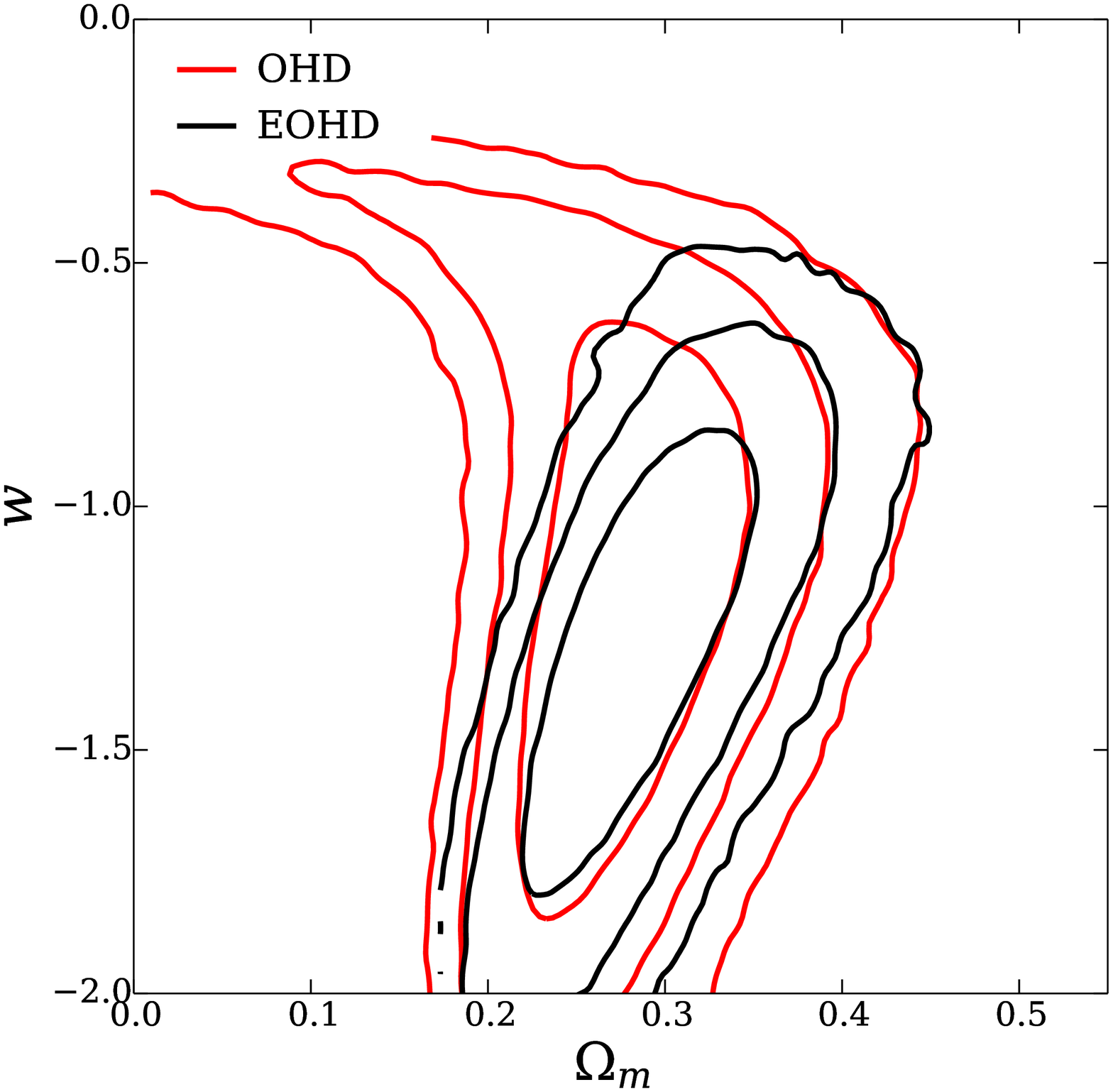}
%\caption{fig1}
\label{fig:side:a}
\end{minipage}%
\begin{minipage}[t]{0.5\linewidth}
\centering
\includegraphics[width=3in]{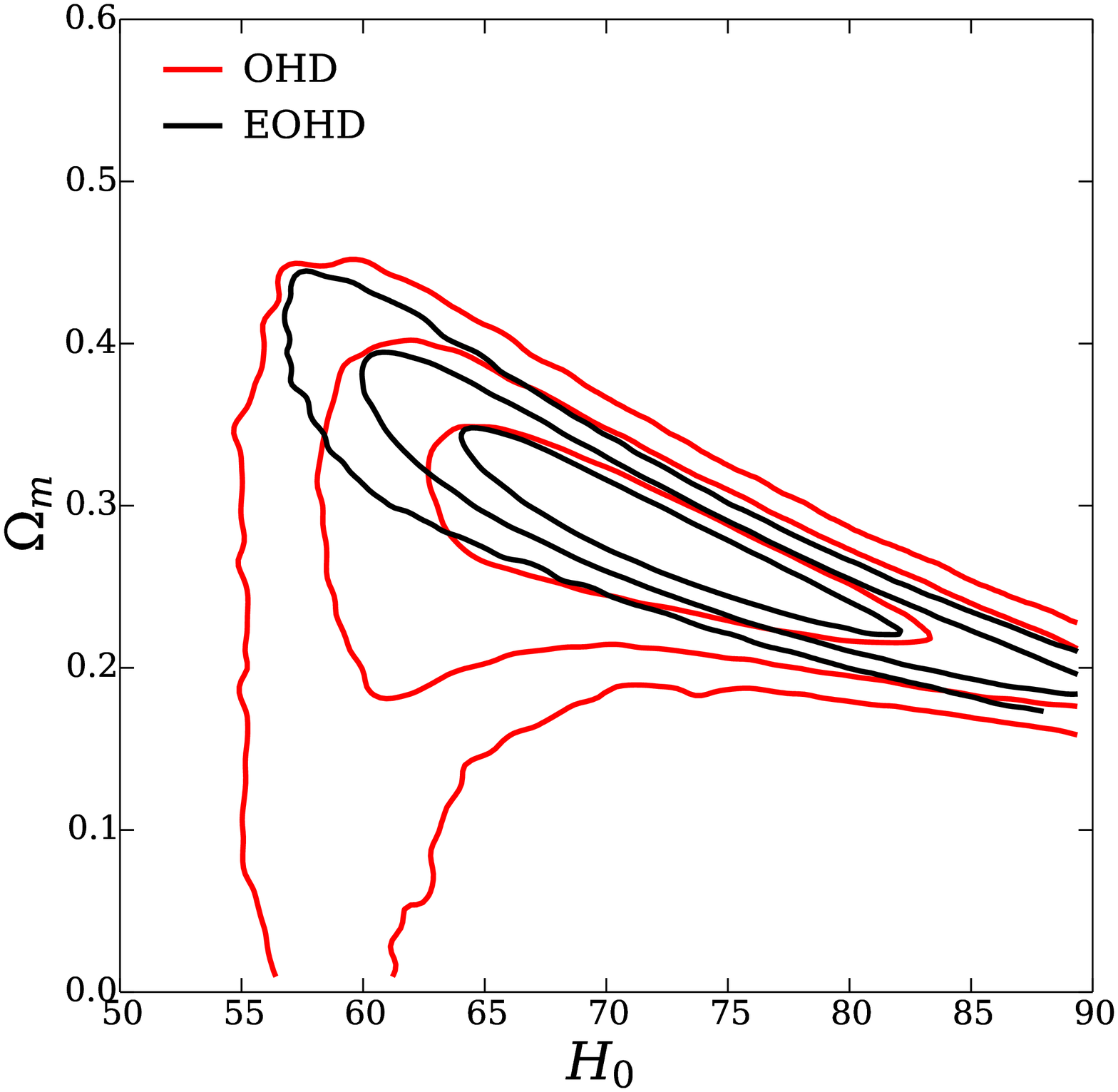}
%\caption{fig2}
\label{fig:side:b}
\end{minipage}
\caption{Constraints on $w$CDM using OHD (red contours) and EOHD (black contours): Marginalized 2-dimensional probability distribution functions for $\Omega_{m}$and $w$, $H_0$ and $\Omega_m$. Compared with the current OHD, the new set perform a better constraint and changes the degeneracy direction slightly.  }
\end{figure}
\begin{table}[h]
\label{table}
\tabcolsep 0pt
\vspace*{-10pt}
\begin{center}
\def\temptablewidth{0.8\textwidth}
{\rule{\temptablewidth}{1pt}}
\begin{tabular*}{\temptablewidth}{@{\extracolsep{\fill}}ccccc}
Model & $\sigma(\boldsymbol{\Omega})$ & OHD & EOHD & $\varepsilon$(\%) \tabularnewline
\hline 
\hline 
 $w$CDM & $\sigma(H_{0})$		 	& 5.84 		& 5.80  	& 0.68  \tabularnewline
 		& $\sigma(\Omega_{m})$ 		& 0.04 		& 0.04 	& -2.2  \tabularnewline
 		& $\sigma(w)$ 				& 0.35  		& 0.30	& 14.28	\tabularnewline
\end{tabular*}
       {\rule{\temptablewidth}{1pt}}
       \end{center}
\vspace*{-18pt}
       \caption{The improvement of EOHD on 1-$\sigma$ error of the cosmological  parameters in the case of $w$CDM. Here $\boldsymbol{\Omega}$ represents the model parameter space. The last column ($\varepsilon$) is the relative improvement ratio for each model parameter defined in Eq.(\ref{31}).}
       \end{table}

\begin{figure}[h]
\label{cov_L}
\begin{minipage}[t]{0.5\linewidth}
\centering
\includegraphics[width=2.9in]{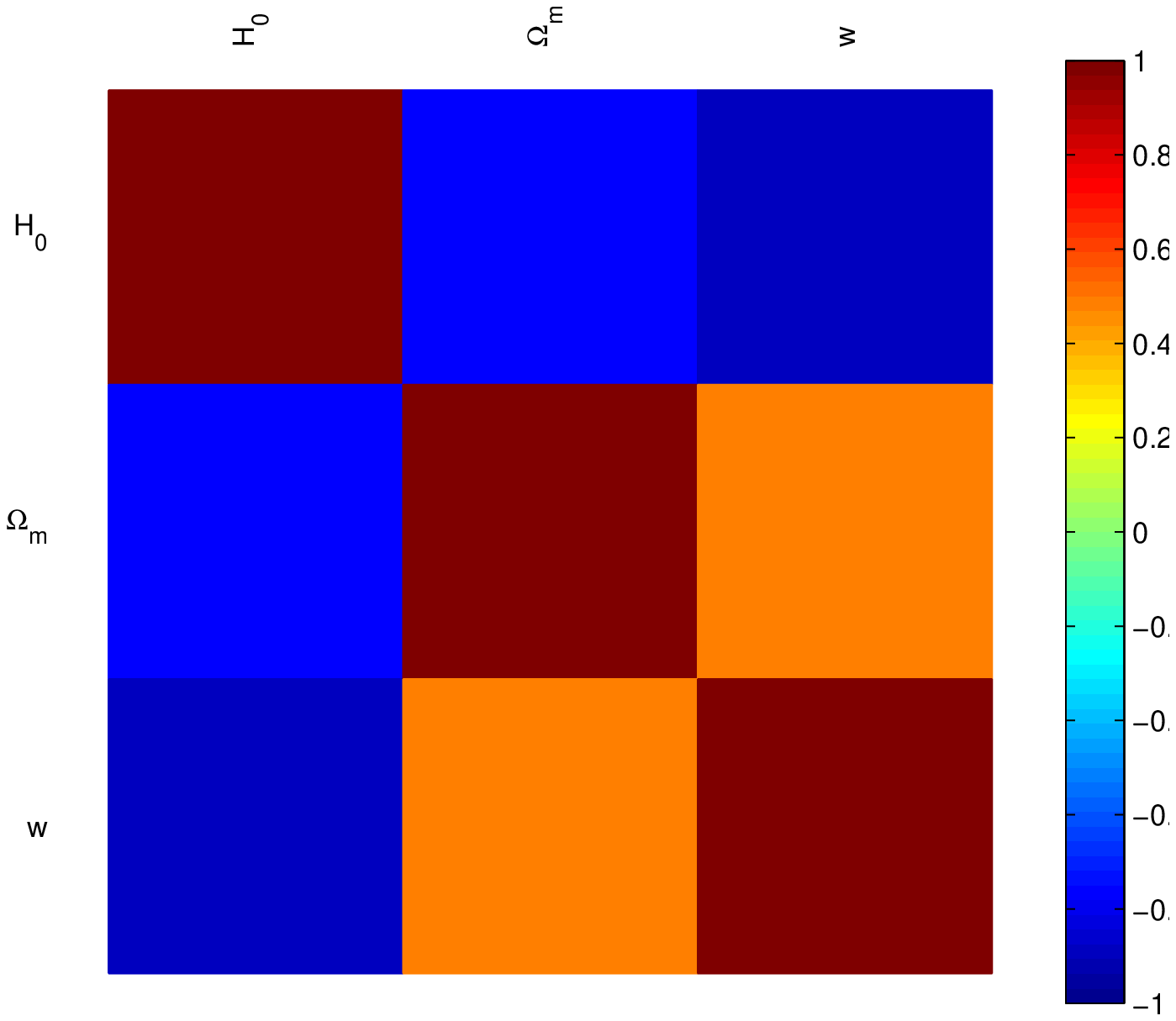}
%\caption{fig1}
\label{fig:side:a}
\end{minipage}%
\begin{minipage}[t]{0.5\linewidth}
\centering
\includegraphics[width=2.9in]{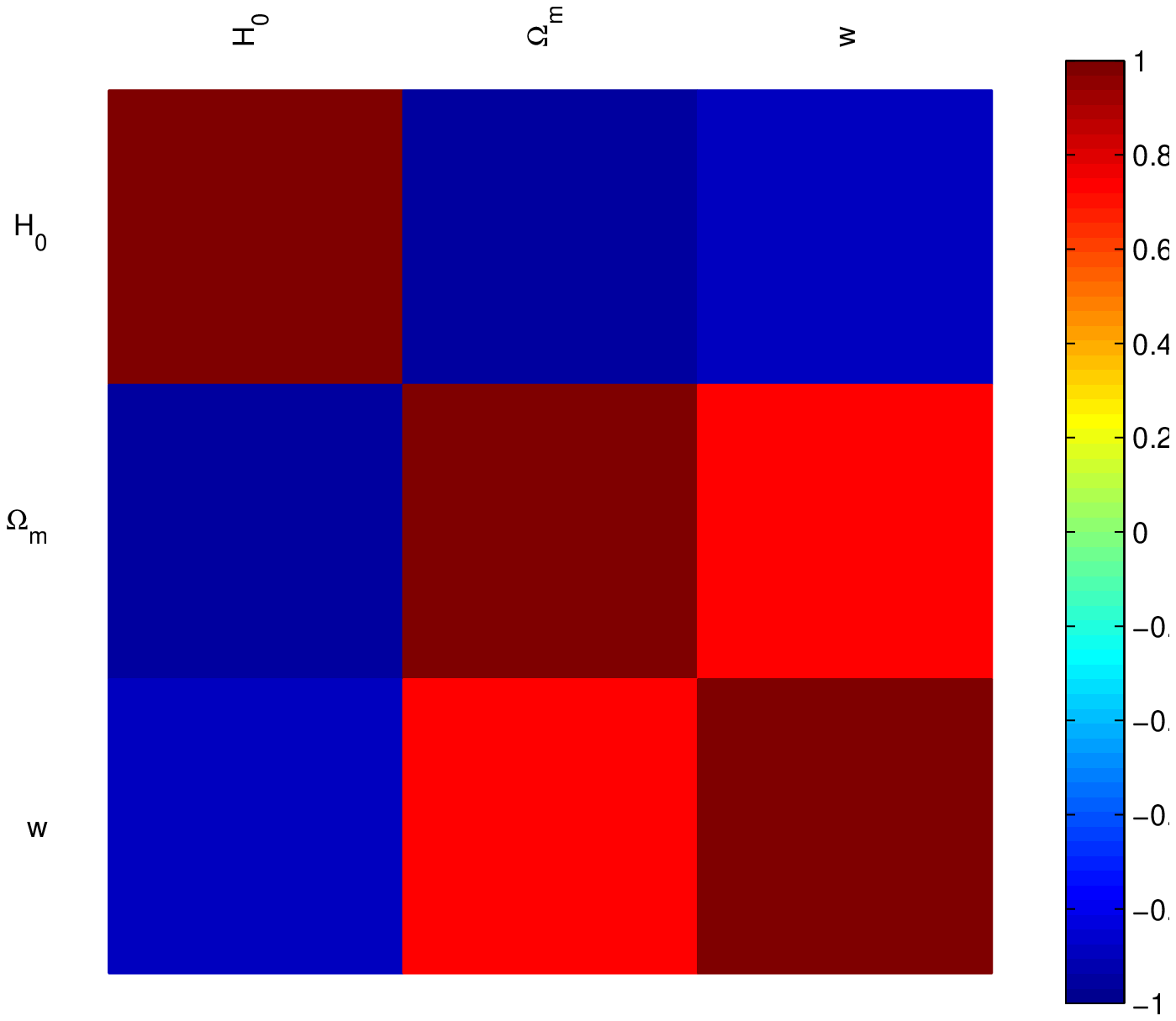}
%\caption{fig2}
\label{fig:side:b}
\end{minipage}
\caption{The correlation matrix $\rho_{ij}$ of $w$CDM. The left side is using OHD and the right one is using EOHD. The two color bars share the same scale.}
\end{figure}

\begin{table}[h]
\label{fom_L_ohd_ohd+}
\tabcolsep 0pt
\vspace*{-10pt}
\begin{center}
\def\temptablewidth{0.8\textwidth}
{\rule{\temptablewidth}{1pt}}
\begin{tabular*}{\temptablewidth}{@{\extracolsep{\fill}}ccccc}
Model & $FoM(\boldsymbol{\Omega}_i,\boldsymbol{\Omega}_j) (i \neq j)$ & OHD & EOHD & $\Delta FoM$(\%) \tabularnewline
\hline 
$w$CDM & $FoM(\Omega_{m},w)$ 			& 98.3	 	& 176 	& 79	 	\tabularnewline
 	   & $FoM(H_0,\Omega_m)$  			& 3.08 		& 3.53 	& 14.6 	\tabularnewline 
\hline 
\end{tabular*}
       {\rule{\temptablewidth}{1pt}}
\end{center}
	   \vspace*{-18pt}
       \caption{The improvement of EOHD on FoM in the case of $w$CDM. Here $\boldsymbol{\Omega}$ represents the model parameter space. The definition of last column is given by Eq.(\ref{fom}).} %The negative larger $\Delta FoM$ is, the 
\end{table}
%------------

\section{The Fisher matrix forecast on  time-dependent $w(z)$}
\label{sec:fisher}
The dark energy is characterized by its equation of state, $w$. Considering a $w$ changing as a function of redshift, $w(z)$, lots of models, such as Quintessence, phantom and Quintoms etc., are used to explain dark energy. Under the phenomenological  parameterization like CPL parameterization $w(z)=w_0+w_a z/(1+z)$ \citep{2001IJMPD..10..213C,2003PhRvL..90i1301L}, the parameter plane $w_0-w_a$ can be divided into four blocks: the Quintom A, Quintom B, Phantom and Quintessence by lines $w_0=1$ and $w_a=-w_0-1 $ (see Fig.1 in \cite{2008IJMPD..17.2025X}). Each of them refers to very different physical background. A question is then raised: can the future CODEX-like survey give $H(z)$ measurement to distinguish these different models? 
\medskip{}

The Fisher matrix forecast is a powerful tool to answer this question. It is an  approximation at first two terms of the likelihood, which is used for designing survey. The Fisher matrix $\mathsf{F}$ in this work is defined as:
\begin{equation}
\label{F}
\mathsf{F}_{ij}= \sum_{k=1}^{N} \frac{\partial{H^{\star}(z_k|\boldsymbol{\Omega})}}{\partial\boldsymbol{\Omega}_i} \frac{1}{\sigma_k^2} \frac{\partial{H^{\star}(z_k|\boldsymbol{\Omega})}}{\partial\boldsymbol{\Omega}_j}
   \end{equation}   
where $H^{\star}(z_k|\boldsymbol{\Omega})$ is the value of OHD for a given cosmology $\boldsymbol{\Omega}=(H_0,\Omega_m,\Omega_\Lambda,w_0,w_a)$:
\begin{equation}
\label{hz2}
H^{\star}(z)=H_{0}\sqrt{\Omega_{m}(1+z)^{3}+\Omega_{\Lambda}(1+z)^{3(1+w(z))}}.
\end{equation}
The derivative values in Eq.(\ref{F}) is obtained at the given fidicual cosmology $\boldsymbol{\Omega}_{\mathrm{fid}}$.
\medskip{}

To test if the SL OHD scheme can distinguish between different models, we  perform the Fisher matrix forecast on these models. Here we take a more ambitious estimation on this future experiment: supposing that all QSO targets from SDSS DR7\footnote{For simplity, we suppose that all of these targets can be obsered by our survey. \url{http://www.sdss.org/dr7/products/value_added/qsocat_dr5.html}} will be put into a 5-year SL signal experiment based on a CODEX-like survey, the number counts is $N_{\mathrm{QSO}}=[1893,1201,1028,421,285,93,32]$ in 7 redshift bins $z_\mathrm{QSO}=[2.0,2.5,3.0,3.5,4.0,4.5,5.0]$. 
%with an equal step of $0.02$. 
 Their errors $\sigma_{H(z)}$ can be obtained via Eq.(\ref{conv}) %using the same fiducial as previous sections. 
We pick up four points from $w_0-w_a$ plane: ($w_0,w_a$)=(-1.5,0.8),(-1,0),(-1.2,-0.2) and (-0.8,0.8), fixing other cosmological parameters to be the same set-up as in previous sections. All of these four points are within the 3-$\sigma$ regions of Plank analysis results \citep{2013arXiv1303.5076P}. The Fisher forecast are evaluated on these 4 models and results are shown in Fig.5.
Comparing the solid ellipses(OHD) and the dotted ones(mix set of OHD and 7 forecast points) in Fig.5, we find that the improvements in the Figure of Merit (FoM) of corresponding models are real, but that are not sufficient to exclude the possibility of alternative dark energy models. Hence, the power to 
discriminate different dark energy models is weak, even considering such an optimistic forecast.

\begin{figure} 
\label{fig5}
\includegraphics[width=1.0\textwidth]{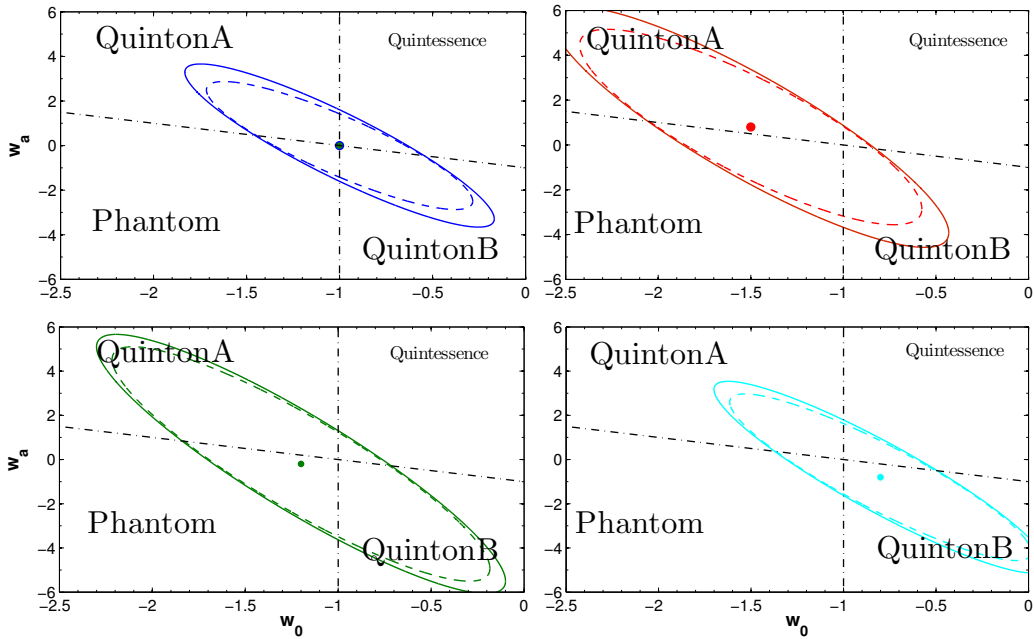}
\caption{Fisher matrix forecast in different fidricual models on $w_0-w_a$ plane. \textit{Leftup:} $(w_0,w_a)=(-1,0)$; \textit{Rightup: }$(w_0,w_a)=(-1,0),$ the Quintom A dark energy; \textit{Leftdown: }$(w_0,w_a)=(-1.2,-0.2)$, the Phontom dark energy; \textit{Rightdown: }$(w_0,w_a)=(-0.8,-0.8)$, the Quintom B dark energy. For each panel, the ellipses in solid line are from the current available OHD and the dotted ones are from the mix set of OHD and optimistic forecast 7 OHD points in Sec.5.2. All the ellipses are correspond to 3$-\sigma$ confidence region.   }
\end{figure}

\section{Conclusion and Discussions}
\label{sec:conclu}
In this paper we propose a new scheme to obtain the 
OHD in the redshift range of $2\lesssim z\lesssim5$ from the Sandage-Loeb signal by observing the Lyman-$\alpha$ forest of QSOs. 
We simulated a CODEX-like survey, estimating new OHD points from SL signal at high redshifts as an addition to the current ones. The mixed data set of the current and the forecast data is called EOHD as a realistic forecast data of SL scheme.  We analyzed the prospects for constraining dark energy models with EOHD, and compare the results with the current OHD.
\medskip{}

The global fitting show that for $\Lambda$CDM, comparing with OHD, all the errors on each single parameter are improved significantly(see Table.2) and the degeneracy between $\Omega_{m}$ and $\Omega_{\Lambda}$ is rotated to the vertical direction of the $\Omega_{m}+\Omega_{\Lambda}=1$ line (see Fig.3). The degeneracy between $H_{0}$ and $\Omega_{m}$ is broken (see Fig.3). 
 In the $w$CDM case, we find that EOHD provides tighter constraints, with an improvement ratio $\varepsilon$ for $w$ of 15.4$\%$.
%the parameter $w$, 
%we found EOHD implies tighter constraints ({\it e.g.} the improvement ratio $\varepsilon$ for $w$ is 15.4$\%$).    
\medskip{}

The capability to distinguish the Quinton A, Quinton B, Phantom and Quintessence models from the $w_0-w_a$ plane under the CPL parameterization is also studied. %We found although the Fisher matrix forecast on SL signal experiment is quite optimistic, it is still hard to distinguish them very well.      
Although the Fisher matrix forecast on the considered SL signal is quite optimistic, we found that it is still hard to distinguish these models clearly. 
\medskip{}

The Sandage-Loeb signal increases the available OHD redshift upper-limit significantly (from $z_\mathrm{max}=2.3$ to $5.0$) and 
%%Different from the former methods, 
%%SL signal provides direct measurement of the cosmic expansion:
 the SL signal scheme is conceptually simple and is a direct probe of cosmic dynamic  expansion, though being observationally challenging. The data processing is also straightforward and there is no need of calibration step like in the case of type  \uppercase \expandafter {\romannumeral 1}a SNe; {\it e.g.} $\Delta v$ obtained from the spectrum flux difference in this experiment between two different observational epochs will reduce the spectrum noise. 
Besides, the errors on data  decrease linearly with observational time interval  $\Delta t_0$ and can be significantly smaller over a few decades. It is hopeful to see small errors in the long enough observational interval.  

\acknowledgments
We are grateful to Yingjie Peng, Hao-Ran Yu, Cong Ma, Feige Wang, and De-Zi Liu for helpful discussion.  We also thank the anonymous referee whose suggestions greatly helped us improve this paper. 
This work was supported by the National Science Foundation of China (Grants No.~11173006), 
the Ministry of Science and Technology National Basic Science program (project 973) under grant No.~2012CB821804, the Fundamental Research Funds for the Central Universities.

\end{document}